\documentclass[aps,twocolumn,showpacs,preprintnumbers,amsmath,amssymb,nofootinbib]{revtex4}
\usepackage{amssymb}
\usepackage{epsfig}
\usepackage{slashed}
\usepackage{graphicx}
\usepackage{bm}
 \usepackage{color}

\begin{document}

\title{Production of doubly heavy-flavored hadrons at $e^+e^-$ colliders}

\author{Xu-Chang Zheng\footnote{e-mail:zhengxc@itp.ac.cn}, Chao-Hsi Chang\footnote{e-mail:zhangzx@itp.ac.cn}, Zan Pan}
\affiliation{ CCAST (World Laboratory),
P.O.Box 8730, Beijing 100190,
China.\\
State Key Laboratory of Theoretical Physics, Institute of Theoretical Physics, Chinese Academy of Sciences, Beijing 100190, China.}

\begin{abstract}
Production of the doubly heavy-flavored hadrons ($B_c$ meson, doubly heavy
baryons $\Xi_{cc}$, $\Xi_{bc}$, $\Xi_{bb}$, their excited states and
antiparticles of them as well) at $e^+e^-$ colliders is investigated under 
two different approaches: $LO$ (leading order QCD complete calculation)
and $LL$ (leading logarithm fragmentation calculation). The results
for the production obtained by the approaches $LO$ and $LL$, including the
angle distributions of the produced hadrons with unpolarized and polarized
incoming beams, the behaviors on the energy fraction
of the produced doubly heavy hadron and comparisons between the two approaches' results,
are presented in terms of tables and figures. Thus characteristics of the production
and uncertainties of the approaches are shown precisely, and it is concluded that only if
the colliders run at the eneries around $Z$-pole (which may be called as $Z$-factories) and 
additionally the luminosity of the colliders is as high as possible then to study
the doubly heavy hadrons thoroughly is accessible. \\

\end{abstract}

\maketitle

\section{Introduction}
\label{intro}

Studies of the production and decays of the doubly heavy flavored hadrons ($B_c$ meson and the baryons $\Xi_{cc}, \Xi_{bc}, \Xi_{bb}$ etc
and their excited states as well as their antiparticle) might be called as `doubly heavy flavored physics'. It is interesting
because the concerned hadrons carry two heavy flavors explicitly and their components move non-relativistically inside the hadrons,
so the `heavy nature' makes perturbative QCD (pQCD) applicable in certain aspects for theoretical predictions, and through their
decays one can study the decays of the heavy flavors inside them etc.

Of them, $B_c$ meson, the unique doubly heavy flavored meson in standard model (SM), was firstly observed in 1998
by CDF collaboration\cite{CDF}, although the observation was confirmed quite soon by themselves and several other
experimental groups such as D0, LHCb and CMS \cite{CDF1, D0, LHCb}; whereas
of the doubly heavy baryons, only $\Xi_{cc}$\footnote{Throughout the paper, $\Xi_{cc}$ denotes $\Xi_{cc}^{+}$ or $\Xi_{cc}^{++}$, $\Xi_{bc}$ denotes $\Xi_{bc}^0$ or $\Xi_{bc}^+$, and $\Xi_{bb}$ denotes $\Xi_{bb}^-$ or $\Xi_{bb}^0$.} was reported to be observed
in a fixed target experiment by SELEX collaboration\cite{selex1,selex2,selex3}, but the SELEX observation has not been confirmed so far,
and there is no experimental report on $\Xi_{bc}$ and $\Xi_{bb}$ at all. The fact shows that experimentally to observe (discovery)
the doubly heavy flavor hadrons is a very tough job. The difficulties for experimental observing the double heavy hadrons
come from two-fold: smallness of the production cross-sections\footnote{The smallness of the production cross-sections causes
that to observe the doubly heavy hadrons, including $B_c$ meson, is accessible now only at high energy hadronic colliders
Tevatron and LHC.} and experimental capabilities in rejecting the backgrounds.

Generally, at a hadronic collider, such as Tevatron or LHC, the production of doubly heavy flavored hadrons is through collision of partons inside the colliding hadrons, so the momentum longitudinal components ($p_\parallel$) of the products in the production and followed decays do not contain ¡®useful¡¯ information, because the center mass system of the colliding partons cannot be well-determined experimentally, so the measured longitudinal component of the momentum of a produced particle always is a combination of the ones of the products in the center mass system and the one of the center mass system itself of the colliding partons. Therefore only perpendicular momenta ($p_\perp$) of the products have substantial meaning (contain useful information). Moreover, in a hadron collision more hadrons, which are not relevant to the concerned parton collision, are produced too, so cause various backgrounds i.e. the environment of the experimental observation is not very clean, thus to make any progress in experimental studies of doubly heavy flavored physics at the environment is comparatively hard. In contrary, the advantages to observe the doubly heavy hadrons in an $e^+e^-$ collider, besides in rejecting the experimental backgrounds, the colliding energy of incoming $e^+,e^-$ is controlled so the longitudinal and perpendicular components, i.e. full momentum, of the produced doubly heavy hadron have substantial meaning well, so the angular distributions of the product too. Thus in this paper we focus the lights on the production of the doubly heavy hadrons in $e^+e^-$ colliders and especially attentions to those ones which run at energies around $Z$-pole (sometimes they are called as $Z$-factories). Note that in reference\cite{eebc1} the authors considered the production of $B_c$ meson etc in an $e^+e^-$ collider too, but they computed only the $p_\perp$ etc. Considering the importance of the characteristics, such as angular distributions etc, of the production, thus here we re-calculate the production and investigate the production under various approaches to implement the defaults of Ref.\cite{eebc1}.

Based on NRQCD\cite{nrqcd}, the production of the doubly heavy hadrons can be estimated quite reliable in term of the factorization into two factors: the involved heavy quark production which is calculable in perturbative QCD, and the relevant matrix elements which
is non-perturbative and to describe the possibility of the produced heavy quarks binding into the doubly heavy hadron. The feature, that the doubly heavy hadrons contain a couple of heavy quarks, and as a core inside the doubly heavy hadrons steadily, may be confirmed in terms of comparisons with available experimental data\cite{ybook}. Only when the doubly heavy hadrons are produced numerously enough, the experimental studies of them may be put into practice. Therefore the theoretical estimates of the production are needed when starting to observe them. Indeed, even theoretical Monte Carlo event generator for the production is available in advance of the experimental observations of $B_c$ meson at Tevatron and LHC.

As well-known, when an $e^+e^-$ collider runs at the energy of $Z$-pole, owing to quite strong resonance effects,
the production cross-sections are quite enhanced\cite{changch}, although the production of the meson $B_c$ is not
enhanced enough, and it was not observed at LEP-I\cite{LEP-I}. Now several $e^+e^-$ colliders with a very high luminosity,
such as ILC, FCC-ee, CEPC and Super Z-factory, are under consideration, and it is planned to run around $Z$-pole energies
for a period at least with a luminosity so greater than LEP-I's by $3-4$ magnitude orders. Thus to re-examine the 
production of the doubly heavy flavored hadrons and the characteristics, especially running at $Z$-pole, 
is freshly interesting. In the present paper we shall estimate the production quantitatively to see the capability 
in studying the doubly heavy flavored hadrons at $e^+e^-$ colliders, and the uncertainties of the theoretical
estimation by comparing the results obtained by the $LO$ and $LL$ approaches precisely.

There is a symmetry between hadrons and anti-hadrons for the production, so the obtained results for
the production of $B_c$ meson, its excited states and doubly heavy baryons are also applicable in the case for their anti-particles.

This paper is organised as follows: In Section II, we present the approaches to compute the doubly heavy hadron production at
$e^+e^-$ colliders. In Section III, we show the numerical results and give a detailed discussion of the results.
Section IV is reserved for discussions and summary.

\section{The approaches for the production}
\label{calctech}

At leading order, the production of $B_c$ meson (its excited states) and doubly heavy
baryons may be computed by two approaches: the so-called QCD complete one ($LO$) and
the fragmentation one ($LL$). Let us apply them to the production and make some comparisons
so as to learn some uncertainties in the estimation of the production.
Firstly we estimate the production by the so-called QCD complete calculation approach, because
from the approach we may extract out the relevant fragmentation functions from an anti-$b$-quark
to $B_c$ meson and from a $c$-quark to $B_c$ meson, which are requested for fragmentation approach.

\subsection{The production of $B_c$ meson and its excited states under the complete QCD approach}
\label{cbcs}

\begin{figure}
\includegraphics[width=0.40\textwidth]{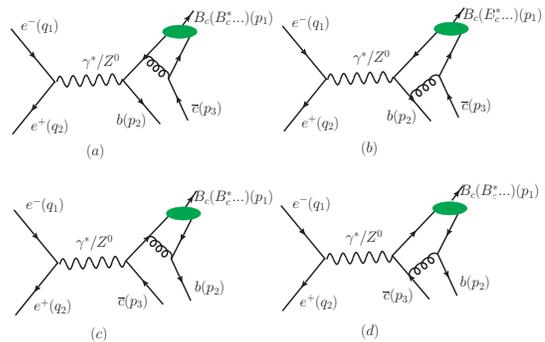}
\caption{Feynman diagrams for the process $e^-(q_1)+e^+(q_2) \rightarrow (c \bar{b})[n](p_1) + b(p_2) +\bar{c}(p_3)$. } \label{feyn}
\end{figure}

The QCD complete calculation for the production of the meson $B_c$ at a $Z$-factory is that
within the framework of NRQCD\cite{nrqcd}, the process
$e^++e^-\rightarrow B_c + b +\bar{c}$ as the production can be factorized into a diquark production of $(c \bar{b})[n]$
i.e. $e^++e^-\rightarrow (c \bar{b})[n] + b +\bar{c}$, and a relevant
nonperturbative matrix element $\langle{\cal O}^H(n)\rangle$ as that in Eq.(\ref{facto}), where $(c \bar{b})[n]$
denotes the diquark with suitable quantum numbers. Then the cross-section for $B_c$ production is follows:
\begin{eqnarray}
\label{facto}
&d\sigma(e^++e^- \to B_c +b+\bar{c})\nonumber \\
&=\sum_{n} d\hat\sigma(e^++e^- \to (c \bar{b})[n]+b+\bar{c})\langle{\cal O}^H(n)\rangle,
\end{eqnarray}
where the nonperturbative matrix element $\langle{\cal O}^{H}(n)\rangle$ (here $H\equiv B_c$) represents
the transition probability from the diquark state $(c\bar{b})[n]$ to the bound state $B_c$. The Feynman
diagrams responsible for the process
$e^++e^- \rightarrow (c \bar{b})[n] + b +\bar{c}$ are described as FIG.\ref{feyn}.
The diquark state $(c\bar{b})[n]$, being color-singlet only\footnote{When considering production of heavy quarkonium ($J/\psi,
\psi', \Upsilon, ......$ etc), the diquark, such as $(c\bar{c})[n]$ or $(b\bar{b})[n]$, is hidden flavored, the color-octet diquark
production can gain some enhancement to compare with color-singlet component, so the color-octet component for hidden flavored
heavy quarkonium production should be considered, although there is some suppression in forming the relevant quarkonium finally.
In contrary to the case of $B_c$ production, the meson $B_c$, so its relevant diquark, carries flavors explicitly, there is no
enhancement effect at all in color octet diquark production, but there still is a strong suppression effect for the
color octet diquark to form $B_c$ meson finally. Therefore to leading order calculations we do not consider the case when the
diquark in color octet at all. Note that according to the `scale rule'\cite{nrqcd}, if the diquark state is color-octet, to compare with the
color-singlet, the relevant matrix element is assumed to be suppressed by $v^2$ (the squared relative velocity of the diquarks inside the doubly heavy hadron).}, is considered here. Moreover, the matrix element can directly relate to the S-wave wave function at origin for $B_c$ meson, and the derivative of the wave function at origin for its P-wave excited states etc under potential framework, and the wave function itself and its derivative at origin can be calculated via potential model \cite{pot1,pot2,pot3,pot4,pot5,pot6}.

It is easy to check that, of the Feynman diagrams FIG.\ref{feyn} responsible for the production,
the first two (a) and (b) are gauge invariant, and the rest two (c) and (d) are also gauge invariant.
Thus it makes sense to separate the contributions from (a) and (b), and from (c) and (d) accordingly.
Furthermore, based on calculations of the contributions from all of the four diagrams, and the contributions
from the two subgroup diagrams mentioned here, one may pick out the contribution from interference of the two
subgroup. In fact, the fragmentation function from an anti-$b$-quark to $B_c$ meson or its excited states is
extracted from the contributions of the subgroup diagrams (a) and (b), and the fragmentation
function from an $c$-quark to $B_c$ meson or its excited states is extracted from the contributions
of the subgroup diagrams (c) and (d).

According to NRQCD, the unpolarized differential cross section $d\hat\sigma(e^++e^-
\to (c \bar{b})[n]+b+\bar{c})$, which stands for the short-distance factor, may be
written down straightforwardly
\begin{equation}
d\hat\sigma(e^++e^- \to (c \bar{b})[n]+b+\bar{c})= \frac{\overline{\sum} |M|^{2} d\Phi_3}{4 \sqrt{(q_1.q_2)^2-m_e^4}} ,
\end{equation}
where $\overline{\sum}$ means that the spins in the initial state are averaged over, and the
spins and colors in final
state are summed over. The three-body phase space for the final state can be represented as
\begin{displaymath}
d{\Phi_3}=(2\pi)^4 \delta^{4}\left(q_1 +q_2 - \sum_{f=1}^3 p_{f}\right)\prod_{f=1}^3 \frac{d^3{p_f}}{(2\pi)^3 2p_f^0}.
\end{displaymath}

The scattering amplitude corresponding to each Feynman diagram
for the process $e^++e^- \rightarrow (c \bar{b})[n] + b +\bar{c}$ can be written as:
\begin{equation}
iM = {\kappa}L_{rr'}^{\mu}D_{\mu\nu}H_{ss'}^{\nu},
\end{equation}
where
\begin{equation}
L_{rr'}^{\mu}={\bar v_r}({q_2}) \Gamma^{\mu}{u_{r'}}({q_1})\label{lepton}
\end{equation}
and
\begin{equation}
H_{ss'}^{\nu}={\bar u_s}({p_2}) {\cal A}^{\nu}{v_{s'}}({p_3})\,.
\end{equation}
For $Z^0$-boson propagation diagrams, here $\kappa=\frac{e^2 g_s^2}{\sin^2\theta_{w} \cos^2\theta_{w}}{\cal{C}}_{ij}$, the color factor
${\cal{C}}_{ij}= \frac{4}{3\sqrt{3}} \delta_{ij}$ (for the diquark $(c\bar{b})[n]$ in color singlet), the vertex $\Gamma^{\mu}=\gamma^{\mu}(1-4\sin^2\theta_w-\gamma_5)$ and $D_{\mu\nu}=\frac{i}{k^2-m_z^2+im_z\Gamma_z}(-g_{\mu\nu}+k_{\mu}k_{\nu}/k^2)$. For  photon propagation diagrams, $\kappa = Q e^2 g_s^2 {\cal{C}}_{ij}$, $\Gamma^{\mu}=\gamma^{\mu}$ and $D_{\mu\nu}=\frac{-ig_{\mu\nu}}{k^2}$, where Q denotes the electric charge of the heavy quark which couples to the virtual photon. ${\cal A}^{\nu}$, corresponding to the precise spin $S$ and orbit
$L$ of the diquak state $(c\bar{b})[n]$, are follows:
\begin{eqnarray}
{\cal A}^{S=0,L=0} &=& {\cal T}\vert_{q=0}, \\
{\cal A}^{S=1,L=0} &=&  \varepsilon_{\beta}^r (p_1)  {\cal T}\vert_{q=0}, \\
{\cal A}^{S=0,L=1} &=& \varepsilon_{\alpha}^l (p_1) \frac{d}{dq_{\alpha}}  {\cal T}\vert_{q=0}, \\
{\cal A}^{S=1,L=1} &=& \varepsilon_{\alpha \beta}^J (p_1) \frac{d}{dq_{\alpha}}  {\cal T} \vert_{q=0} ,
\end{eqnarray}
where $\varepsilon_{\beta}^r (p_1)$, $\varepsilon_{\beta}^l (p_1)$ are the polar vectors relating to spin or orbit angular momentum in triplet
for $(c\bar{b})[n]$-diquark, and $\varepsilon_{\alpha \beta}^J (p_1)\, \;(J=0,1,2)$ is the polar tensor of the diquark in P-wave and spin-triplet states. ${\cal T}_i\, (i=1,2,3,4)$ relating to the Feynman diagrams accordingly in FIG.\ref{feyn} are as follows:
\begin{eqnarray}
{\cal T}_1 &=&  {{\gamma^{\nu}}{\Gamma_{b}} \frac{\slashed{p}_2 - \slashed{k} + {m_b}}{(p_2 - k)^2 - m_b^2}{\gamma^\rho} \frac{\Pi_{p_1}(q)} {(p_{11} + {p_3})^2}{\gamma_\rho}}, \\
{\cal T}_2 &=&  {{\gamma^\rho} \frac{\slashed{k} - \slashed{p}_{12} + {m_b}}{(k - p_{12})^2 - m_b^2} {\gamma^{\nu}}{\Gamma_{b}} \frac{\Pi_{p_1}(q)} {(p_{11} + {p_3})^2}{\gamma_\rho}}, \\
{\cal T}_3 &=&  {{\gamma^\rho} \frac{\Pi_{p_1}(q)} {(p_{12} + {p_2})^2}{\gamma_\rho}} \frac{\slashed{k} - \slashed{p}_3 + {m_c}}{(k - p_3)^2 - m_c^2} {\gamma^{\nu}}{\Gamma_{c}} , \\
{\cal T}_4 &=&  {{\gamma^\rho} \frac{\Pi_{p_1}(q)} {(p_{12} + {p_2})^2} {\gamma^{\nu}}{\Gamma_{c}} \frac{\slashed{p}_{11} - \slashed{k} + {m_c}}{(p_{11} - k)^2 - m_c^2} {\gamma_\rho}},
\end{eqnarray}
where $\Gamma_{c}=\frac{1}{4}-\frac{2}{3}\sin^2\theta_w-\frac{1}{4}\gamma^5$ for $Z-c\bar{c}$ vertex, and $\Gamma_{b}=-\frac{1}{4}+\frac{1}{3}\sin^2\theta_w+\frac{1}{4}\gamma^5$ for $Z-b\bar{b}$ vertex for $Z^0$ propagation
diagrams, and $\Gamma_{b}=\Gamma_{c}=1$ for $\gamma-b\bar{b}$ and $\gamma-c\bar{c}$ vertex for photon propagation diagrams.
$q$ is the relative momentum between the two quarks inside the $(c\bar{b})[n]$-diquark, and $p_{11}$, $p_{12}$ introduced here
are the momenta of the two quarks inside the diquark $(c\bar{b})[n]$, i.e.
\begin{equation}
p_{11}=\frac{m_c}{M}p_1+q ~~{\rm and}~~ p_{12}=\frac{m_b}{M}p_1-q,
\end{equation}
where $M \simeq m_b+m_c$ stands for the mass of $B_c$ meson or its excited states i.e. $(c\bar{b})[n]$ diquark.
The projectors for spin singlet and triplet may be approximately written as follows:
\begin{eqnarray}
\Pi^0_{p_1}(q)= \frac{-\sqrt{M}}{{4{m_b}{m_c}}}(\slashed{p}_{12}- m_b) \gamma_5 (\slashed{p}_{11} + m_c),\\
\Pi^{\beta}_{p_1}(q)= \frac{-\sqrt{M}}{{4{m_b}{m_c}}}(\slashed{p}_{12}- m_b) \gamma^{\beta} (\slashed{p}_{11} + m_c).
\end{eqnarray}
After substituting these projectors into the above amplitudes, then we square the amplitudes, and sum over the final degrees of freedom. For appropriate total angular momentum, we also perform polarization summation properly. We define:
\begin{equation}
\Pi_{\alpha \beta}=-g_{\alpha \beta}+\frac{p_{1\alpha}{p_{2\beta}}}{M^2},
\end{equation}
and the summation over the polarization for a spin triplet S-wave state or a spin singlet P-wave state is given by
\begin{equation}
\sum\limits_{J_z}\varepsilon_{\alpha} \varepsilon^*_{\alpha'}=\Pi_{\alpha \alpha'},
\end{equation}
where $J_z$ can be $r_z$ or $l_z$ for spin and orbit projection respectively. As for the three
multiplets $^3P_J$ with J=0, 1 and 2, the summation over the polarizations is given by
\begin{eqnarray}\label{3pja}
\varepsilon^{(0)}_{\alpha\beta} \varepsilon^{(0)*}_{\alpha'\beta'} &=& \frac{1}{3} \Pi_{\alpha\beta}\Pi_{\alpha'\beta'}, \\
\sum_{J_z}\varepsilon^{(1)}_{\alpha\beta} \varepsilon^{(1)*}_{\alpha'\beta'} &=& \frac{1}{2}
(\Pi_{\alpha\alpha'}\Pi_{\beta\beta'}- \Pi_{\alpha\beta'}\Pi_{\alpha'\beta}), \label{3pjb}\\
\sum_{J_z}\varepsilon^{(2)}_{\alpha\beta} \varepsilon^{(2)*}_{\alpha'\beta'} &=& \frac{1}{2}
(\Pi_{\alpha\alpha'}\Pi_{\beta\beta'}+ \Pi_{\alpha\beta'}\Pi_{\alpha'\beta})-\frac{1}{3}
\Pi_{\alpha\beta}\Pi_{\alpha'\beta'} . \label{3pjc} \nonumber \\
\end{eqnarray}

Since technically the colliding $e^+e^-$ beams for a $Z$-factory can be polarized, so it is interesting to consider the cases.
To calculate the polarized cross sections for polarized electron and positron beams, we choose the the helicity states of
the initial electron and positron to describe their polarization states, and we can use the projection operators
\begin{equation}
P_L=\frac{1-\gamma^5}{2},  ~~ P_{R}=\frac{1+\gamma^5}{2},
\end{equation}
to project out the polarized cross section. We note that in our calculations, the masses of electron and positron are ignored,
so under the approximation and with a gauge boson photon or $Z$ boson mediating as shown in FIG.\ref{feyn},
only two polarization cases, $e_R^+ e_L^-$ and $e_L^+ e_R^-$, contributes to the production.
Thus now the polarized cross sections can be obtained by means of substituting Eq.(\ref{lepton}) with
\begin{equation}
{\bar v_r}({q_2}) \Gamma^{\mu}P_L{u_{r'}}({q_1}),
\end{equation}
and
\begin{equation}
{\bar v_r}({q_2}) \Gamma^{\mu}P_R{u_{r'}}({q_1}),
\end{equation}
for $e_R^+ e_L^-$ and $e_L^+ e_R^-$, respectively.

Putting on the nonperturbative matrix
element $\langle{\cal O}^H(n)\rangle$ ($H\equiv B_c$) accordingly as that in Eq.(\ref{facto}), the cross sections for $B_c$-meson
production are obtained by integrating over the final phase space. For the
production of excited states, the cross sections can be obtained similarly, as long as
$[n]$ in $(c\bar{b})[n]$ and $H$ in $\langle{\cal O}^H(n)\rangle$ are matched properly.

\subsection{The fragmentation functions and production under the fragmentation approach}

The fragmentation approach for the production is based on the perturbative QCD factorization
theorem: to calculate the relevant quark-pair production processes $e^++e^-\rightarrow
b + \bar{b}$ and $e^++e^-\rightarrow c + \bar{c}$ respectively first and then to apply the
fragmentation functions for a $\bar{b}$-quark to $B_c$
meson and for a $c$-quark to $B_c$ meson accordingly to the quark-pair production processes at the factorization
energy scale $\mu$ properly. Here, more precisely, the fragmentation functions for the production are extracted from a relevant process
under QCD complete calculations as the above, then need further to do evolution of them to the factorization energy scale $\mu$.

Namely the inclusive differential cross section can be written as:
\begin{eqnarray}\label{eq101}
&\displaystyle \frac{d\sigma(z,\mu)}{dz}=\sum_i \int_z^1 \frac{dy}{y}\times \nonumber\\
&\displaystyle \frac{d\sigma(e^++e^-\to i(y)+X,\mu)}{dy} D_{i\to B_c}(z/y,\mu)\,,
\end{eqnarray}
where the $i$ represents the mediate partons, i.e. $c$-quark or $\bar{b}$-quark, and
$z$ represents the energy fraction of the $B_c$-meson from $i$-quark, here it can be defined as
\begin{equation}
z=\frac{2k\cdot p_1}{k^2},
\end{equation}
and $k$ is the four-momentum of the $Z$-boson or the photon, $\mu$ is the factorization
energy-scale. Up-to LO and at the energy scale $\mu\simeq m_{_Z}$, the factorization formula reads
\begin{equation}\label{eq102}
\frac{d\sigma}{dz}=\sum_i \sigma^{LO}(e^++e^-\rightarrow i+\bar{i}) D_{i\to B_c(B^*_c,\cdots)}(z,\mu),
\end{equation}
here the mediate parton $i$ can be either $c$-quark or $\bar{b}$-quark. Note that all
the large logarithm terms $ln(m_{_Z}^2/\mu^2)$, which come from the collinear emissions
for high order calculations, also disappear due to the factorization energy scale
$\mu$ being set to $m_{_Z}$.

The strategy for the fragmentation approach is firstly to extract the fragmentation functions
from the process $e^-+e^+\to b+\bar{c}+B_c(B^*_c,\cdots)$ near the threshold $k^2=4(m_c+m_b)^2$
i.e. the fragmentation functions $D_{\bar b\to B_c(B^*_c,\cdots)}(z,\mu_0)$ with $\mu_0\simeq m_b+2m_c$
and $D_{c\to B_c(B^*_c,\cdots)}(z,\mu_0)$ with $\mu_0 \simeq m_c+2m_b$ are obtained by the
extraction, then to do evolution on them to the fragmentation energy scale $\mu=m_{_Z}$, and
finally the production we are interested in here is computed out in terms of
Eq.(\ref{eq102}). Now let us show how to extract the fragmentation functions from the perturbative calculations
and then to do the evolution to the factorization energy scale $\mu=m_{_Z}$ for application here.

Precisely we only describe the way for extracting the fragmentation function of $B_c$-meson, because
extracting the fragmentation functions of the excited states is similar, as long as
the different quantum numbers of the diquark $(c\bar{b})[n]$ and different matrix element $\langle{\cal O}^H(n)\rangle$
are taken into account.

For convenience, we define some kinematic variables as follows:
\begin{equation}
x=\frac{2k\cdot p_2}{k^2},\, y=\frac{2k\cdot p_3}{k^2},\, d=\frac{M^2}{k^2},\,  r_1=\frac{m_c}{M},\, r_2=\frac{m_b}{M}\,,
\end{equation}
here $M=m_{_{B_c}}$ when $B_c$ fragmentation function is concerned, and $k$ is the momentum of the virtual photon or $Z$-boson in FIG.\ref{feyn}.

\begin{widetext}

To extract $D_{\bar{b}\to B_c}(z,\mu_0)$, the differential cross sections
and total cross sections of the production $e^++e^-\to b+\bar{c}+B_c$ at $k^2\geq 4(m_c+m_b)^2$
via $\bar{b}$-quark production by mediating virtual photon (here $k^2$ is much smaller than $m^2_Z$, so
the contributions from $Z$-boson mediation can be ignored) i.e. of diagrams (a) and (b) in FIG.\ref{feyn}
only those via virtual photon are needed to be computed precisely\footnote{For extracting the fragmentation functions (e.g. $D_{\bar{b}\to B_c}(z,\mu_0)$), to do the LO calculations is suitable at the collision energy just above that a little higher than the threshold of the relevant quark pair production.}. Namely the fragmentation functions $D_{\bar{b}\to B_c}(z,\mu_0)$ with $\mu_0\simeq m_b+2m_c$ is extracted with the process $e^++e^-\to b+\bar{c}+B_c$
just at the threshold (above a little) as
\begin{eqnarray}
D_{\bar{b}\to B_c}(z,\mu_0)=\frac{1}{\sigma^{LO}(e^+e^-\to b\bar{b})}\frac{d\sigma(e^+e^-\to\bar{b}\to B_c)}{dz}\,.
\end{eqnarray}
If $d\ll 1$, the double differential cross section of S-wave states can be
written approximately as
\begin{eqnarray}\label{eq0001}
&\displaystyle\frac{1}{\sigma_0} \frac{d^2 \sigma}{dx~ dz} &\displaystyle\simeq \frac{8 \alpha_s^2(4m^2_c)
\vert \Psi_S(0) \vert^2 }{  27 k^2 M r_1^2 (1-r_2 z)^2} \displaystyle \left[ \frac{w_0(z)}{(1-x)^2} + \frac{d~w_1(z)}{(1-x)^3} + \frac{d^2 ~w_2(z)}{(1-x)^4}\right]\,,
\end{eqnarray}
here $\sigma_0=\sigma^{LO}(e^++e^-\rightarrow b+\bar{b})$. As pointed out by the authors of Ref.\cite{zbc1},
owing to $d\ll 1$, the terms with positive and larger than -2 power of (1-x) may be ignored, but when x being integrated,
a factor $1/d$ is obtained form each $1/(1-x)$ factor, so the terms containing
$w_0(z)$, $w_1(z)$ and $w_2(z)$ in Eq.(\ref{eq0001}) should be kept.
The integral bounds of x at a fixed z now can be written as
\begin{eqnarray}
(1-x)_{min} \simeq  \frac{(1-r_2 z)^2 d}{z(1-z)}\,,\;\;\;
(1-x)_{max} \simeq  z-\frac{(1-r_1 z)^2 d}{z(1-z)}\,,
\end{eqnarray}
and the integral over the range of $x$ is performed, so the fragmentation functions are obtained:
\begin{eqnarray}
D_{\bar{b}\to B_c(B_c^*)}(z,\mu_0)= \frac{8 \alpha_s^2(4m^2_c) \vert \Psi_S(0) \vert^2 }{  27 m_c^2 M (1-r_2 z)^2}
\left[\delta w_0(z) + \frac{\delta^2}{2} w_1(z)+ \frac{\delta^3}{3} w_2(z)\right]\,,
\end{eqnarray}
where $\delta=\frac{z(1-z)}{(1-r_2 z)^2}$.
When the fragmentation functions for P-wave states $B_{c,J(P)}^*$, i.e. $(\bar{b}c)$ in P-wave states, are extracted,
the relevant double differential cross sections are replaced as
\begin{eqnarray}
\frac{1}{\sigma_0} \frac{d^2 \sigma}{dx~ dz} = \frac{128 \alpha_s^2(4m^2_c) \vert \Psi'_P(0) \vert^2 }{  27 k^2 M^3 r_1^4 (1-r_2 z)^4}
\left[ \frac{w_0(z)}{(1-x)^2} + \frac{d~w_1(z)}{(1-x)^3} + \frac{d^2 ~w_2(z)}{(1-x)^4} +
\frac{d^3 ~w_3(z)}{(1-x)^5} + \frac{d^4 ~w_4(z)}{(1-x)^5} \right] ,
\end{eqnarray}
where the coefficients $w_0(z)$, $w_1(z)$, $w_2(z)$, $w_3(z)$ and $w_4(z)$ for P-wave states can be computed\cite{zbc1,fragbc1,fragbc2}. Integrating the double differential cross sections over variable $x$, we obtain
\begin{eqnarray}
&\displaystyle D_{\bar{b}\to B_{c,J(P)}^*}(z,\mu_0)= \frac{128 \alpha_s^2(4m^2_c) \vert \Psi'_P(0) \vert^2 }{27 m_c^4 M (1-r_2 z)^4}
&\left[\delta w_0(z) + \frac{\delta^2}{2} w_1(z)+ \frac{\delta^3}{3} w_2(z)+ \frac{\delta^4}{4} w_3(z)+ \frac{\delta^5}{5} w_4(z)\right] .
\end{eqnarray}

For extracting the fragmentation functions $D_{c\to B_c(B^*_c,\cdots)}(z,\mu_0)$,
the way is similar but the diagrams (c) and (d) instead of the diagrams (a) and (b) in FIG.\ref{feyn}
are computed. One may also find that by interchanging $r_1 \leftrightarrow r_2$ and $\alpha_s^2(4m^2_c)$,
$\mu_0=2m_c+m_b$ being replaced by $\alpha_s^2(4m^2_b)$, $\mu_0=m_c+2m_b$ respectively, so
the fragmentation functions $D_{c\to B_c(B^*_c,\cdots)}(z,\mu_0)$ may be obtained in terms of the
interchanges from the one $D_{\bar{b}\to B_c(B^*_c,\cdots)}(z,\mu_0)$ as pointed.

\end{widetext}

Having the fragmentation functions, we need to do the needed evolution from $\mu_0$ to the energy scale $\mu=m_{_Z}$, where
the factorization is made for the concerned production.

The evolution of the fragmentation functions are governed by the DGLAP equation \cite{AP},
\begin{eqnarray}
&&\mu \frac{\partial}{\partial\mu}D_{i\to B_c(B^*_c,\cdots)}(z,\mu) \nonumber\\
&=&\frac{\alpha_s(\mu)}{\pi} \sum_j \int^1_z \frac{dy}{y}P_{i\to j}(z/y)D_{j\to B_c(B^*_c,\cdots)}(y,\mu),
\end{eqnarray}
where $i$ and $j$ are all possible partons, but here $j=i$.
For $B_c(B^*_c,\cdots)$-meson production, $i$ can be $c$-quark or $\bar{b}$-quark, and
\begin{equation}
P_{i\to i}(z)=\frac{4}{3}\left( \frac{1+z^2}{1-z} \right)_+ ,
\end{equation}
is the splitting function. Using the approximate method to solve the DGLAP equation \cite{APsolve}, we obtain
\begin{eqnarray}
&&D_{i\to B_c(B^*_c,\cdots)}(z,\mu)=R \otimes D_{i\to B_c(B^*_c,\cdots)}(z,\mu_0)\nonumber \\
&&+ \kappa R \otimes P_{\Delta} \otimes D_{i\to B_c(B^*_c,\cdots)}(z,\mu_0),
\end{eqnarray}
where $\kappa = \frac{2}{\beta_0}{\rm ln}\left( \frac{\alpha_s(\mu_0^2)}{\alpha_s(\mu^2)}\right) $, it is not the coefficient $\kappa$ we defined in the amplitudes in Sec.\ref{cbcs}, $\otimes$ is the convolution which is defined as
\begin{equation}
A \otimes B=\int^1_z \frac{dy}{y}A(z/y)B(y),
\end{equation}
and
\begin{equation}
R(z)=\frac{\left(-{\rm ln}(z)\right)^{8\kappa/3-1}}{\Gamma(8\kappa/3)},
\end{equation}
and
\begin{equation}
P_{\Delta}(y)=\frac{4}{3}\left[ \frac{1+y^2}{1-y}+\frac{2}{{\rm ln}(y)} +(\frac{3}{2}-2\gamma_E)\delta(1-y) \right].
\end{equation}
Thus the fragmentation functions are evolved to the energy scale of $m_{_Z}$ with the formulas.


\subsection{The production of doubly heavy baryons}

In this subsection we consider the production of the doubly heavy baryons $(QQ'q)$, where $Q$ and $Q'$ mean heavy quarks
$c$ and/or $b$ but $q$ means light quark.

The approach, which we adopt here, for the production is to divide it into two steps. The first step is that a binding
diquark $<QQ'>$ ($Q$ and $Q'$ may $b$-quark or $c$-quark) with `suitable' quantum numbers is produced, precisely
$<QQ'>$ may be in color anti-triplet or sextuplet, and the second step is that the diquark 'fragments' to a doubly
heavy baryon $(QQ'q)$ through caching a light quark from 'environment' with a fragmentation probability of
almost one hundred percent, i.e. it is implied that the produced doubly heavy diquark always makes a doubly heavy baryon accordingly.
Here for the first step, we compute the production of doubly heavy diquarks under the QCD complete approach similar to that
of $B_c$ meson but the matrix element $\langle{\cal O}^{B_c}(n)\rangle$ should be replaced by a proper one $\langle{\cal O}^{<QQ'>}(n)\rangle$. Namely
\begin{eqnarray}
&d\sigma(e^++e^- \to <QQ'> +\bar{Q}+\bar{Q'})=\nonumber \\
&\sum_{n} d\hat\sigma(e^++e^- \to (QQ')[n]+\bar{Q}+\bar{Q'})\times\nonumber\\
&\langle{\cal O}^{<QQ'>)}(n)\rangle\,.
\end{eqnarray}

We should note
here that the energy scale in fragmentation approach is $2m_Q+m_{Q'}$ or $m_Q+2m_{Q'}$ similar to the case of $B_c$ meson.

For the production of $<bc>$-diquark, there are four Feynman diagrams responsible for it, which are the same as that of $B_c$ meson production showed in FIG.\ref{feyn} except the fermion line of the b-quark being reversed. For the production of $<cc>$-diquark and $<bb>$-diquark, besides the four diagrams corresponding to those showed in FIG.\ref{feyn}, there are more four diagrams obtained by an interchange of two quark lines of the diquark.

In order to calculate the amplitude of the diquark production, we adopt the method used in Refs.\cite{eeXi,doublybaryon3}. We reverse one of the fermion lines of the heavy quark through charge conjugate transformation firstly, and then similarly to writing down the amplitude of doubly heavy meson production showed above, we directly write down the requested amplitude for the binding diquark production. Precisely a fermion line appearing in diquark production can be written as,
\begin{eqnarray}
Am=\bar{u}_{s_1}\left( \frac{m_Q}{M}p_1\right)\Gamma_n s_f(k_{n-1},m_Q)\times \nonumber\\
\cdots s_f(k_1,m_Q)\Gamma_1 v_{s_2}(p_2),
\end{eqnarray}
where $s_f(k_i,m_Q)$ is the fermion propagator, $\Gamma_i$ is the interaction vertex, $m_Q$ is heavy quark mass of the fermion line, and $n$ is the number of the interaction vertices. Using the charge conjugation matrix $C=-i\gamma^2\gamma^0$, we can obtain the following transformation,
\begin{eqnarray}
&& v^T_{s_2}(p_2)C = -\bar{u}_{s_2}(p_2), C^-\bar{u}_{s_1}\left( \frac{m_Q}{M}p_1\right)^T=v_{s_1}\left( \frac{m_Q}{M}p_1\right),\nonumber \\
&& C^-(\gamma^{\alpha})^TC = -\gamma^{\alpha}, ~~~~C^-(\gamma^{\alpha}\gamma^{5})^TC = \gamma^{\alpha}\gamma^{5} \nonumber \\
&& C^-s_f^T(-k_i,m_Q)C = s_f(k_i,m_Q)\,. \label{CTran}
\end{eqnarray}
If the fermion line contains no axial vector vertex, we have
\begin{widetext}
\begin{eqnarray}
Am^T &=&v_{s_2}^T(p_2) \Gamma_1^T s_f^T(k_1,m_Q) \cdots s_f^T(k_{n-1},m_Q) \Gamma_n^T \bar{u}_{s_1}\left( \frac{m_Q}{M}p_1\right)^T\nonumber \\
&=&v_{s_2}^T(p_2) C C^-\Gamma_1^T C C^- s_f^T(k_1,m_Q)C \cdots C^- s_f^T(k_{n-1},m_Q) C C^- \Gamma_n^T C C^- \bar{u}_{s_1}\left( \frac{m_Q}{M}p_1\right)^T\nonumber \\
&=&(-1)^{(n+1)}\bar{u}_{s_2}(p_2) \Gamma_1  s_f(-k_1,m_Q) \cdots s_f(-k_{n-1},m_Q) \Gamma_n v_{s_1}\left( \frac{m_Q}{M}p_1\right)\,.\label{CTran2}
\end{eqnarray}

\end{widetext}
If the fermion line contains an axial vector vertex, then a (-1) factor should be contained. After reversing the heavy qaurk line, we can transform the amplitude of the diquark production to meson production except an additional $(-1)^{(n+1)}$ factor for pure vector case ($(-1)^{(n+2)}$ factor for containing an axial vector case). i.e. the amplitude of $(bc)[n]$-diquark production can be represented as
\begin{equation}
M_{diquark}=(M^a_1-M^v_1)+(M^a_2-M^v_2)+M_3+M_4,
\end{equation}
where $M_1, M_2, M_3, M_4$ are the amplitude of $B_c$ production showed above, and $M^a_i$,$M^v_i$ are those for the cases containing an axial vector or pure vector cases of $M_i$ respectively. As we have mentioned above, for the $<cc>$-diquark and $<bb>$-diquark productions there are four more diagrams come from the exchange of the two identical quark line inside the diquark. Since the relative velocity between the two quarks inside the diquark, $v$, is set to be zero up-to the lowest order non-relativistic approximation, the contributions from additional diagrams are the same as the four diagrams relating to heavy quarkonium production. There is one (1/2!) factor comes from wave function of identical particles in diquark. Thus we only need to calculate the four diagrams, and multiplying the cross section by $(2^2/2!)=2$.

According to the decomposition $\textbf{3}\otimes\textbf{3}=\bar{\textbf{3}}\oplus\textbf{6}$, $(bc)[n]$-diquark can be in $\bar{\textbf{3}}$ or $\textbf{6}$ color state. It is known that `one-gluon exchange' interaction is important inside the diquark, and the interaction inside the diquark  in $\bar{\textbf{3}}$ state is attractive, but that in $\textbf{6}$ state is repulsive, so people believe that the 'binding force' (including the confinement interaction besides one-gluon exchange) in $\bar{\textbf{3}}$ is stronger than that in $\textbf{6}$, i.e. we have the consequence  $\langle{\cal O}^{<QQ'>}\rangle\, _{\bar{\textbf{3}}}>\langle{\cal O}^{<QQ'>}\rangle\, _{\textbf{6}}$, hence in this paper we will ignore the contributions from the color $\textbf{6}$ component. According to the request of Fermi-Dirac statistics for quarks, the wave functions of the diquark should be totally antisymmetric under exchange of the two quarks inside the diquarks, so for the production of the ground state of the diquaks, $<bb>$ and $<cc>$, which is concerned here only, the diquark wave function should be S-wave and in color anti-triplet and spin triplet.

The color factor for color anti-triplet diquark production can be represented as follows:
\begin{equation}
{\cal{C}}_{ij}={\cal{N}}\sum_{a}\sum_{m,n} T^a_{im}T^a_{jn}\varepsilon_{mnk},
\end{equation}
where ${\cal{N}}=1/ \sqrt{2}$ is the normalization constant, $\varepsilon_{mnk}$ is the the antisymmetric tensor for the $\bar{\textbf{3}}$ state and it is satiesfy $\varepsilon_{mnk}\varepsilon_{m'n'k}=\delta_{mm'}\delta_{nn'}-\delta_{mn'}\delta_{nm'}$. After squaring we obtain that ${\cal{C}}^2_{ij}=4/3$ for the $\bar{\textbf{3}}$ state.

\section{Numerical Results}
\label{numer}

\subsection{The production of $B_c(B^*_c,\cdots)$-mesons}

When doing the numerical calculations under complete QCD approach and fragmentation approach,
the input parameters are taken as follows:
\begin{eqnarray}
& &m_b=4.9 ~{\rm GeV}, m_c=1.5 ~{\rm GeV}, m_Z=91.1876~ {\rm GeV},\nonumber \\
& & \sin^2 \theta_w=0.231, \alpha=1/128,\vert R_S(0) \vert^2=1.642 ~{\rm GeV^3},\nonumber \\
& &\vert R'_P(0) \vert^2=0.201~ {\rm GeV^5},
\end{eqnarray}
where $\alpha=\alpha(m^2_{_Z})$ is the electromagnetic coupling constant; $ R_S(0)$ and $R'_P(0)$ are the ridial wave function of the
($B_c(B^*_c,\cdots)$) and its derivative of the relevant states at origion respectively and taken from potential model \cite{pot6}.

\begin{widetext}

As for the strong coupling constant, we take $\alpha_s(m^2_Z)=0.1185$ \cite{PDG} as the reference point, and by one loop running coupling
we obtain the request values $\alpha_s(4m^2_c)=0.237$ and $\alpha_s(4m^2_b)=0.175$.

\begin{table}[h]
\begin{tabular}{c c c c c}
\hline\hline
~contribution~  & ~total~ & ~$\bar{b}$-frag.~ & ~$c$-frag.~ & ~interference~  \\
\hline
$\sigma(B_c,\;^1S_0)$ & 2.734 & 2.613 & 5.20$\times10^{-2}$ & 6.90$\times10^{-2}$ \\
$\sigma(B^*_c,\;^3S_1)$ & 3.823 & 3.722 & 4.45$\times10^{-2}$ & 5.65$\times10^{-2}$ \\
$\sigma(B^{**}_c,\;^1P_1)$ & 0.271 & 0.269 & 3.01$\times10^{-3}$ & -1.01$\times10^{-3}$ \\
$\sigma(B^{**}_c,\;^3P_0)$ & 0.164 & 0.157 & 8.13$\times10^{-3}$ & -1.13$\times10^{-3}$ \\
$\sigma(B^{**}_c,\;^3P_1)$ & 0.340 & 0.331 & 5.77$\times10^{-3}$ & 3.23$\times10^{-3}$ \\
$\sigma(B^{**}_c,\;^3P_2)$ & 0.365 & 0.366 & 3.87$\times10^{-4}$ & -1.39$\times10^{-3}$ \\
\hline\hline
\end{tabular}
\caption{The cross section (in $pb$) of $e^-e^+ \to B_c(B^*_c,\cdots)+b+\bar{c}$ at $Z$-pole under complete QCD calculation. $\bar{b}$-frag., $c$-frag. and 'Interferance' denote the contributions from the first two diagrams in FIG.\ref{feyn}, those from the last two diagrams in FIG.\ref{feyn} and those from the interference of the first two diagrams and the last two diagrams in FIG.\ref{feyn}.}
\label{bcsection}
\end{table}

The two diagrams (a), (b) and the other two diagrams (c), (d) as well of FIG.\ref{feyn} are gauge invariant, thus in complete QCD calculation
approach to separate the contributions from diagrams (a), (b) and those from diagrams (c), (d) as well as those from their 'interference' makes
sense. Especially note that in fragmentation approach the fragmentation from $\bar{b}$-quark takes into account the contribution from
diagrams (a), (b) and the fragmentation from $c$-quark takes into account the contribution from diagrams (c), (d) respectively, whereas
the interference contributions cannot be taken into account.


The numerical results of the cross sections at $Z$-pole under the QCD cmplete calculation with separation of the contributions:
from (a), (b) diagrams ($\bar{b}$-frag.), from (c), (d) diagrams ($c$-frag.) and from interference are put
in TABLE.\ref{bcsection} \footnote{Our results are slightly different from Ref.\cite{eebc1}, the main reason is that we adopted different values for the electomagnetic coupling constant and the strong coupling constant, another reason is that Ref.\cite{eebc1} put an additional (-1) factor in $Z-b\bar{b}$ interaction vertex. If we change these differences to the same as Ref.\cite{eebc1}, we obtain the same results.}.

\begin{table}[h]
\begin{tabular}{c c c c}
\hline\hline
~contribution~  & ~total~ & ~$\bar{b}$-frag.~ & ~$c$-frag.~  \\
\hline
$\sigma(B_c,\;^1S_0)$ & 2.850 & 2.792 & 5.88$\times10^{-2}$  \\
$\sigma(B^*_c,\;^3S_1)$ & 3.974 & 3.923 & 5.08$\times10^{-2}$  \\
$\sigma(B^{**}_c,\;^1P_1)$  & 0.295 & 0.292 & 3.53$\times10^{-3}$  \\
$\sigma(B^{**}_c,\;^3P_0)$ & 0.170 & 0.160 & 9.08$\times10^{-3}$  \\
$\sigma(B^{**}_c,\;^3P_1)$  & 0.360 & 0.354 & 6.59$\times10^{-3}$  \\
$\sigma(B^{**}_c,\;^3P_2)$  & 0.396 & 0.395 & 4.74$\times10^{-4}$  \\
\hline\hline
\end{tabular}
\caption{The cross section (in $pb$) of $e^-e^+ \to  B_c(B^*_c,\cdots)+\cdots$ at the $Z$-pole under fragmentation approach.}
\label{bcsecfrag}
\end{table}

For comparison between the complete QCD approach (LO) and the fragmentation approach (LL), the results under the fragmentation approach are put in TABLE.\ref{bcsecfrag}, where the contributions from $\bar{b}$-quark fragmentation and $c$-quark fragmentation are showed explicitly. One may see that the results for total cross-sections obtained from the two approaches are consistent each other. This is because that the property $\int_0^1 dz P_{i\to i}(z)=0$ leads to non-evolution with $\mu$ for $\int_0^1 dz D_{i\to B_c(B^*_c,\cdots)}(z,\mu)$\cite{fragbc1}, so the total cross sections are same as the ones at the initial scale. It is easy to see that the fragmentation contributions dominantly come from the $\bar{b}$-quark fragmentation, i.e. the contributions from diagrams (a) and (b) in FIG.\ref{feyn} under the complete QCD calculation. The contributions from $c$-quark fragmentation are comparatively small and it is due to deeper depress from the gluon propagator in diagrams (c) and (d) than from that in diagrams (a) and (b).


The collision energy of initial $e^+e^-$ may not exactly at $Z$-pole, so we also show the behave of the cross sections around the $Z$-pole. We present the cross sections at several collison energies around the $Z$-pole in TABLE.\ref{bcEcm}, and from it one can see that the cross sections decrease rapidly when the collision energy deviate from $m_Z$. When the collision energy deviate from $m_Z$ by $5$ GeV, i.e. about twice width of the $Z^0$ boson, the cross sections depressed by about one order, thus when studying $B_c$ and its excited states, sometimes it is crucial to make use of the effects of $Z^0$ resonance at an $e^+e^-$ collider.


\begin{table}
\begin{tabular}{c c c c c c c c c c c c c c}
\hline\hline
~~$(\sqrt{s}-m_{_Z})$({\rm GeV})~~  & ~~-5~~ & ~~-2.5~~ & ~~-1.5~~ & ~~-0.8~~   & ~~-0.4~~   & ~~-0.2~~  & ~~0~~    & ~~0.2~~   & ~~0.4~~   & ~~0.8~~ & ~~1.5~~ & ~~2.5~~ & ~~5~~ \\
\hline
$\sigma(B_c,\;^1S_0)$  & 0.15  & 0.53   & 1.09 & 1.91 &2.46  &2.65 &2.73 & 2.68 & 2.50 & 1.97 & 1.15 & 0.56 & 0.17   \\
$\sigma(B^*_c,\;^3S_1)$ & 0.21  & 0.74  & 1.52 & 2.67 & 3.44  & 3.71 &3.82  & 3.74  & 3.50  & 2.75  & 1.60  & 0.79  & 0.24 \\
$\sigma(B^{**}_c,\;^1P_1)$  & 0.01   & 0.05    & 0.11   & 0.19  & 0.24  & 0.26 &0.27   & 0.27   & 0.25   & 0.19  & 0.11  & 0.06  & 0.02\\
$\sigma(B^{**}_c,\;^3P_0)$ & 0.01   & 0.03    & 0.07   & 0.11  & 0.15  & 0.16 &0.16   & 0.16  & 0.15  & 0.12  & 0.07  & 0.03  & 0.01  \\
$\sigma(B^{**}_c,\;^3P_1)$ & 0.02   & 0.07   & 0.14  & 0.24  & 0.31  & 0.33 &0.34   & 0.33  & 0.31  & 0.24   & 0.14  & 0.07  & 0.02 \\
$\sigma(B^{**}_c,\;^3P_2)$ & 0.02   & 0.07   & 0.15  & 0.25  & 0.33  & 0.35 &0.37  & 0.36 & 0.33 & 0.26 & 0.15  & 0.08 & 0.02  \\
\hline\hline
\end{tabular}
\caption{The cross section (in $pb$) of $e^-e^+ \to B_c(B^*_c,\cdots)+b+\bar{c}$ varies with the collision energy around $m_{_Z}$ under complete QCD approach.}
\label{bcEcm}
\end{table}
\vspace{8mm}

\end{widetext}

\begin{figure}[!h]
\includegraphics[width=0.45\textwidth]{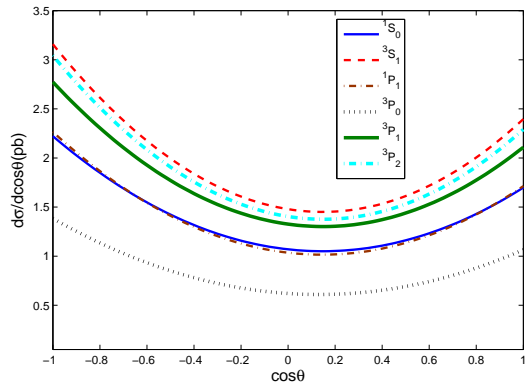}
\caption{The differential cross sections $d\sigma/d\cos{\theta}$ for the production $e^+e^- \to B_c(B^*_c,\cdots) +b+\bar c$
under complete QCD approach. In order to put all the results into the figure, the results of P-wave meson production are multiplied by a factor ten.} \label{bccos}
\end{figure}

\begin{figure}[!h]
\includegraphics[width=0.45\textwidth]{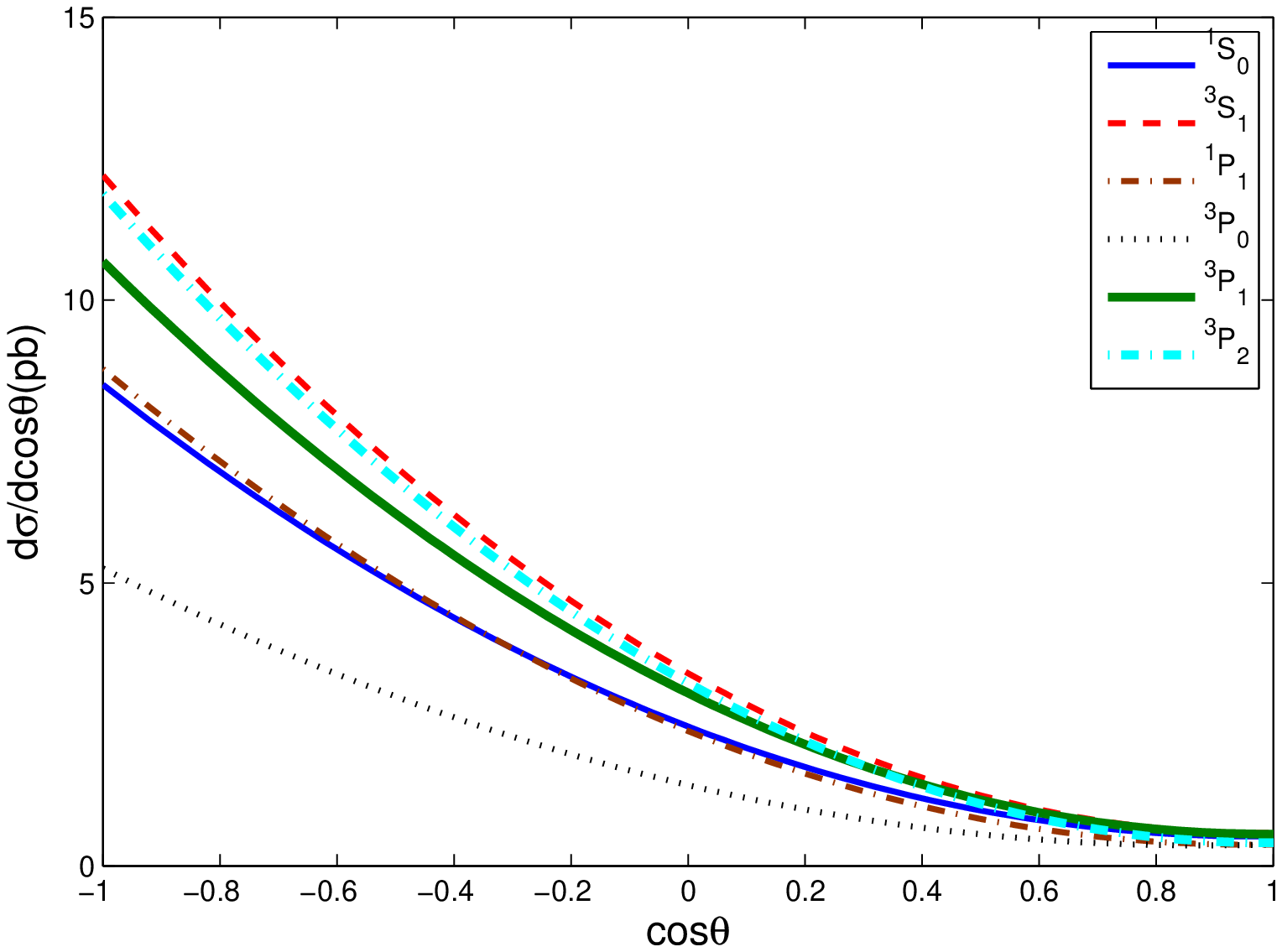}
\caption{The differential cross sections $d\sigma/d\cos{\theta}$ for the production $e_R^+e_L^- \to B_c(B^*_c,\cdots) +b+\bar c$  (with polarized
incoming beams) under complete QCD approach.  In order to put all the results into the figure the results for P-wave states are multiplied by a factor ten.} \label{bccosl}
\end{figure}

\begin{figure}[!h]
\includegraphics[width=0.45\textwidth]{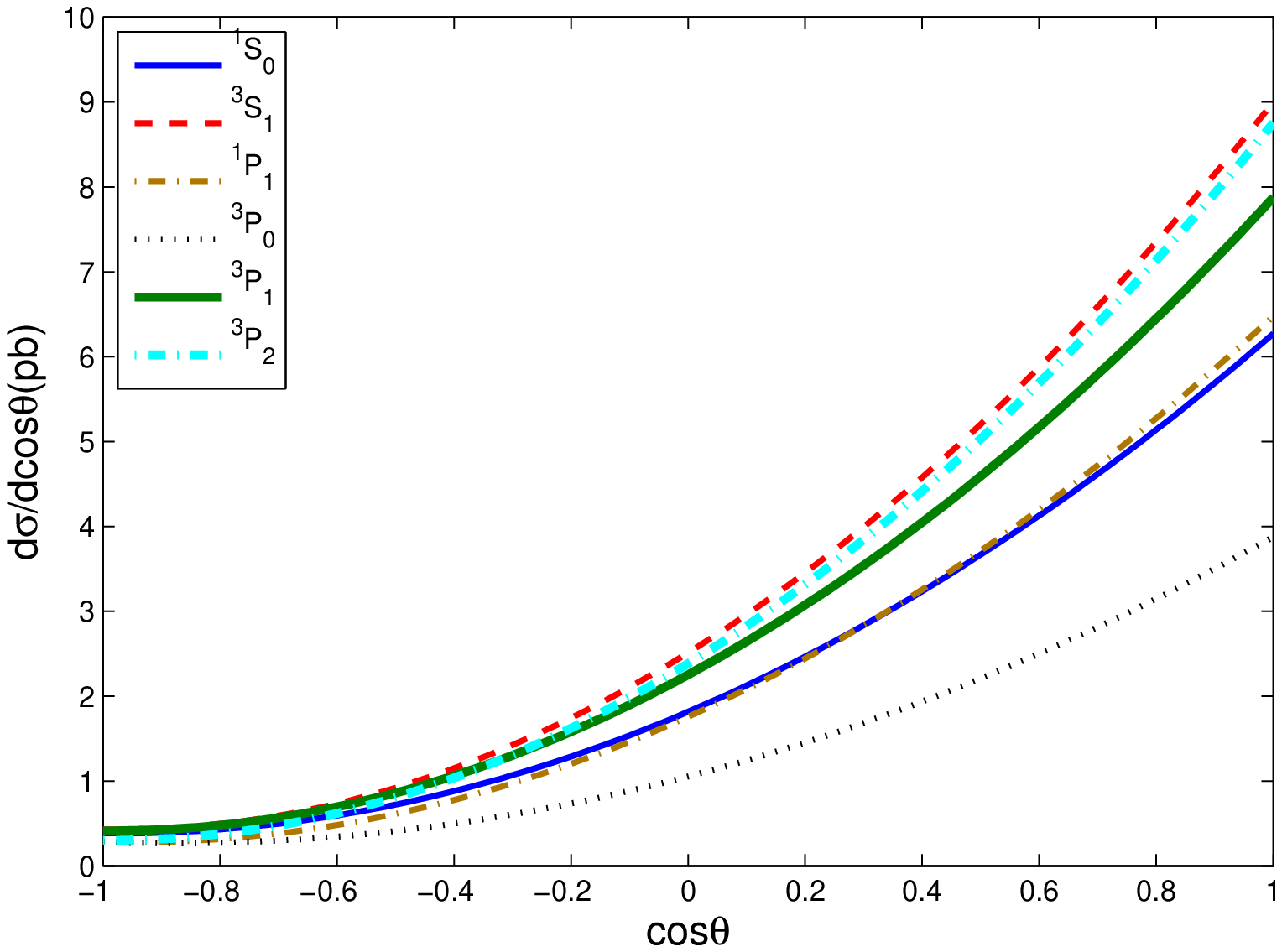}
\caption{The differential cross sections $d\sigma/d\cos{\theta}$ for the production $e_L^+e_R^- \to B_c(B^*_c,\cdots) +b+\bar c$ (with polarized
incoming beams) under complete QCD approach.  In order to put all the results into the figure the results for P-wave states are multiplied by a factor ten.} \label{bccosr}
\end{figure}

\begin{figure}[!h]
\centering
\includegraphics[width=0.45\textwidth]{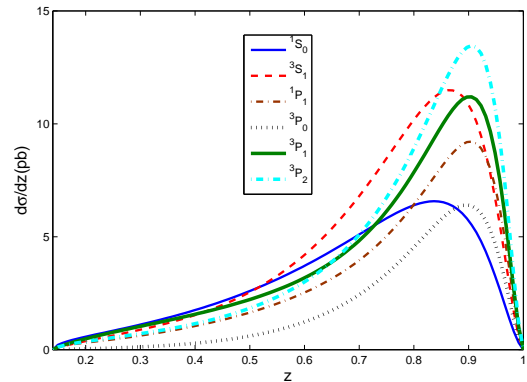}
\caption{The differential cross sections $d\sigma/dz$ for the processes $e^+e^- \to B_c(B^*_c,\cdots) +b+\bar c$ by complete QCD approach,
where $z$ is the energy fraction carried by the concerned meson. The results for P-wave states are multiplied by a factor ten.} \label{bcz1}
\end{figure}
\begin{figure}[!h]
\centering
\includegraphics[width=0.45\textwidth]{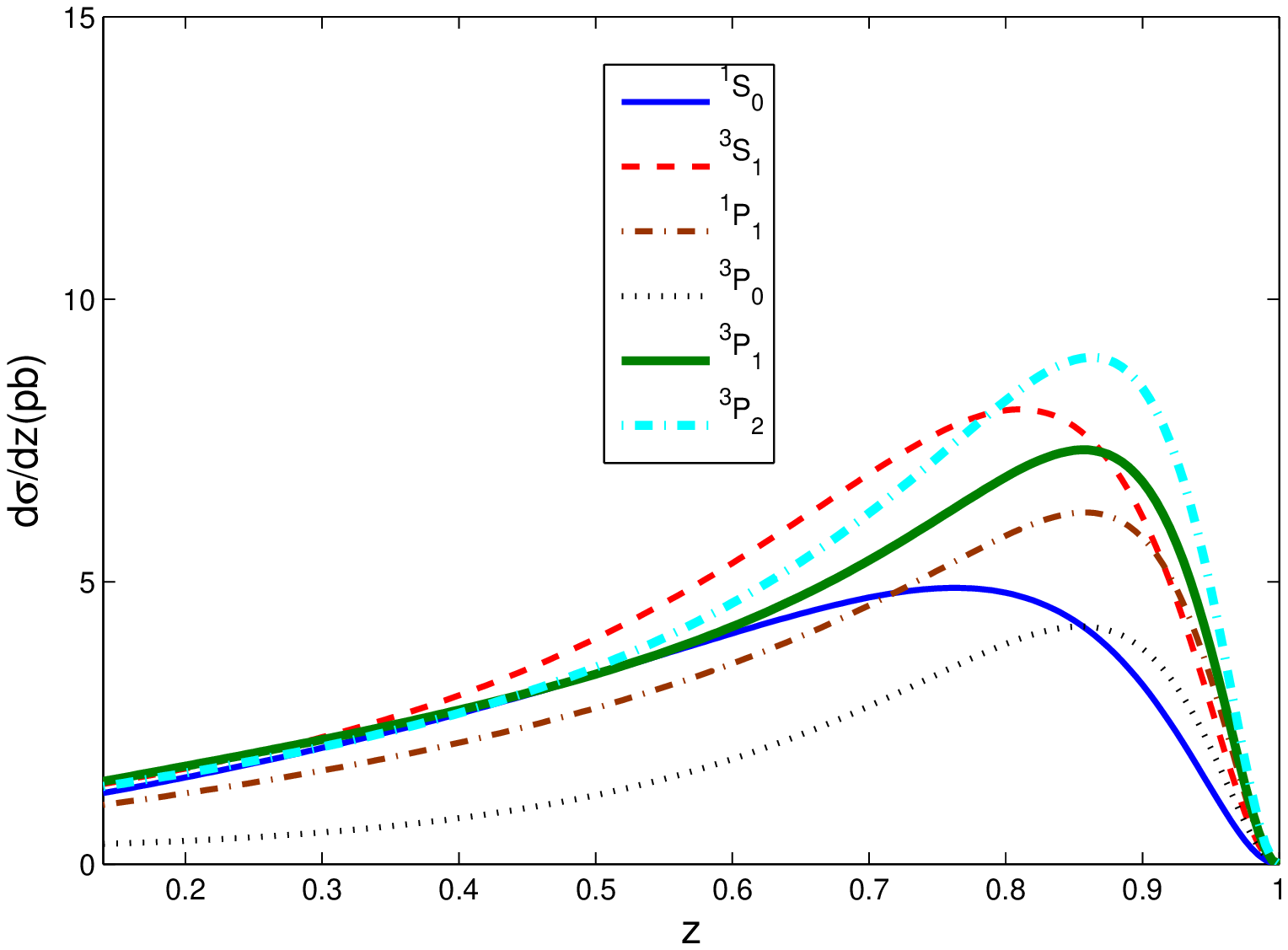}
\caption{The differential cross sections $d\sigma/dz$ for the processes $e^+e^- \to B_c(B^*_c,\cdots) +\cdots$ by $\bar{b}$-quark fragmentation
under fragmentation approach, where $z$ is the energy fraction carried by the concerned meson. The results for P-wave states are multiplied by a factor ten.}
\label{bcz2}
\end{figure}
\begin{figure}[!h]
\centering
\includegraphics[width=0.45\textwidth]{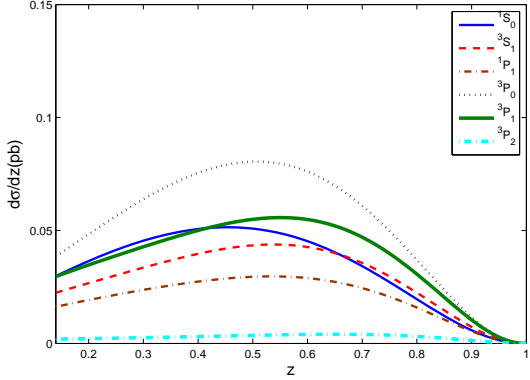}
\caption{The differential cross sections $d\sigma/dz$ for the processes $e^+e^- \to B_c(B^*_c,\cdots) +\cdots$ by $c$-quark fragmentation
under fragmentation approach, where $z$ is the energy fraction carried by the concerned meson. The results for P-wave states are multiplied by a factor ten.}
\label{bcz3}
\end{figure}


In order to know the characteristic of the production for experimental studying the mesons, we reserve the numerical results for the differential cross sections $d\sigma/d\cos{\theta}$ where $\theta$ is the angle between the momenta of the meson in final state and the electron in initial state
under complete QCD approach, and present them in FIG.\ref{bccos}.

From FIG.\ref{bccos}, one may see clearly that the production has maximum near the collision axis. The differential cross sections also show us certain backward-forward asymmetry in the direction of the initial positron. It is caused by parity-violation in the interaction mediated by $Z^0$ boson exchange.

Since the incoming beams for the production may be polarized, so we also compute and present the angle distributions for polarized electron and and polarized positron beams in FIGs.\ref{bccosl},\ref{bccosr}. But note that just for convenience, here we take helicities to take into account the polarizations. One may see the difference in magnitude between $Z^0$ boson coupling to left-handed helicity and right-handed helicity electron, and this leads the backward-forward asymmetry in FIG.\ref{bccos}.

In FIG.\ref{bcz1}, we present the energy fractions ($z$ dependence) of $B_c(B^*_c,\cdots)$ meson obtained by complete QCD approach (LO). The results are obtained by $\bar{b}$-quark fragmentation and by $c$-quark fragmentation under fragmentation approach are presented in FIG.\ref{bcz2}
and in FIG.\ref{bcz3}. Here
for the fragmentation approach, we have evolved the scale of the fragmentation functions to $m_Z$. From the figures one may see that the contributions from $c$-quark fragmentation are smaller than those from $\bar{b}$-quark fragmentation by two magnitude orders, thus the $c$-quark fragmentation contributions in the production can be ignored. From FIGs.\ref{bcz1},\ref{bcz2} one may see that the production of $B_c(B^*_c,\cdots)$ meson at low energy fraction (small $z$) estimated by the fragmentation approach is greater than that by complete QCD approach. Namely from the figures, one can see the differences from each other in distributions by the approaches clearly.

\begin{table}[h]
\begin{tabular}{c c c}
\hline\hline
~$\sqrt{s}$(GeV)~  & ~180~ & ~240~   \\
\hline
$\sigma(B_c,\;^1S_0)$ & 1.05 & 0.47  \\
$\sigma(B^*_c,\;^3S_1)$  & 1.55 & 0.72  \\
$\sigma(B^{**}_c,\;^1P_1)$  & 0.11 & 0.05  \\
$\sigma(B^{**}_c,\;^3P_0)$ & 0.07 & 0.03  \\
$\sigma(B^{**}_c,\;^3P_1)$  & 0.14 & 0.07  \\
$\sigma(B^{**}_c,\;^3P_2)$  & 0.15 & 0.07  \\
\hline\hline
\end{tabular}
\caption{The cross sections (in $fb$) of the production $e^-e^+ \to B_c(B^*_c,\cdots)+b+\bar{c}$ at 180 GeV and 240 GeV under complete QCD approach.}
\label{bccepc}
\end{table}

An $e^++e^-$ collider running at energies 180 GeV and 240 GeV (above the threshold of WW and/or ZH production respectively) is
very suitable for studying the properties of W-boson and Higgs boson etc, so to see the possibility for experimental studying $B_c(B^*_c,\cdots)$
mesons at the collider at the same time is interesting. Therefore, we also estimate the
meson production at the high energies for an $e^++e^-$ collider, and put the results in TABLE.\ref{bccepc}. The results show that the cross sections
are in $fb$, less than the cross sections at the $Z$-pole by $3\sim4$ order, that means it is hopeless for studying $B_c(B^*_c,\cdots)$
meson at such an $e^++e^-$ collider running at the high energies even having so high luminosity as $\sim 10^{36}cm^{-2}s^{-1}$.

\subsection{The production of doubly heavy baryons}

To do the numerical calculations on the production of doubly heavy baryons, we adopt the parameters as follows,
\begin{eqnarray}
&&m_b=5.1 ~{\rm GeV}\,,\;\; m_c=1.8 ~{\rm GeV}\,, \nonumber\\
&&\vert \Psi_{(cc)}(0) \vert^2=0.039 ~{\rm GeV^3}\,,\;\;\vert \Psi_{(bc)}(0) \vert^2=0.065 ~{\rm GeV^3}, \nonumber \\
&&\vert \Psi_{(bb)}(0) \vert^2=0.152 ~{\rm GeV^3}\,,
\end{eqnarray}
which are the smae as Ref.\cite{doublybaryon4}. Note here that for the production of $\Xi_{cc}$ and $\Xi_{bc}$, the squared momentum
transferred $Q^2$ carried by gluon is adopted as $4m^2_c$, whereas for the $\Xi_{bb}$, it is adopted as $4m^2_b$.

The cross sections of the diquark production obtained by our estimation:
\begin{eqnarray}
&&\sigma_{\vert\left( cc \right){\scriptsize _{\overline{\textbf{3}}}}\,, ^3S_1 \rangle}=0.52~ {\rm pb},~~
\sigma_{\vert\left( bc \right){\scriptsize _{\overline{\textbf{3}}}} \,, ^1S_0 \rangle}=0.58 ~{\rm pb},\nonumber \\
&&\sigma_{\vert\left( bc \right){\scriptsize _{\overline{\textbf{3}}}}\,, ^3S_1 \rangle}=0.79~ {\rm pb},~~
\sigma_{\vert\left( bb \right){\scriptsize _{\overline{\textbf{3}}}} \,, ^3S_1 \rangle}=0.04 ~ {\rm pb}.\nonumber
\end{eqnarray}

\begin{figure}[!h]
\centering
\includegraphics[width=0.45\textwidth]{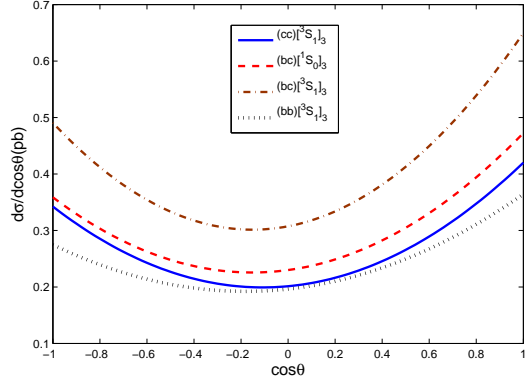}
\caption{Differential cross sections $d\sigma/d\cos{\theta}$ for the production of the doubly heavy baryons under complete QCD approach, where $\theta$ is the angle between the momenta of the final baryon and the initial electron. In order to show in one figure, for $\Xi_{bb}$, it was multiplied by a factor ten.} \label{Xicos}
\end{figure}

For the doubly heavy baryon production, we present the angle distributions (differential cross sections) in FIG.\ref{Xicos} too, where the distributions for various channels are labeled precisely. One may see from the figure that the doubly heavy baryon production has certain backward-forward asymmetry too. However, the distributions are different from those of the $B_c$ meson production, namely the doubly heavy baryons are mainly concentrated in the direction of the initial electron. One can understand the reason by the fragmentation approach, that the mainly contribution of the $B_c$-meson production comes from the diagrams (a) and (b) in FIG.\ref{feyn}, i.e. from the fragmentation of $\bar{b}$-quark. Since the produced $\bar{b}$-quark is concentrated in the direction of the initial positron around $Z$-pole, so there are more $B_c$ mesons along the direction of the initial positron. In contrast, the $\Xi_{cc}$ mainly from the fragmentation of c-quark, $\Xi_{bc}$ and $\Xi_{bb}$ from b-quark, where b-quarks and c-quarks are mainly along the direction of intial electron around $Z$-pole.

\begin{figure}[!h]
\centering
\includegraphics[width=0.45\textwidth]{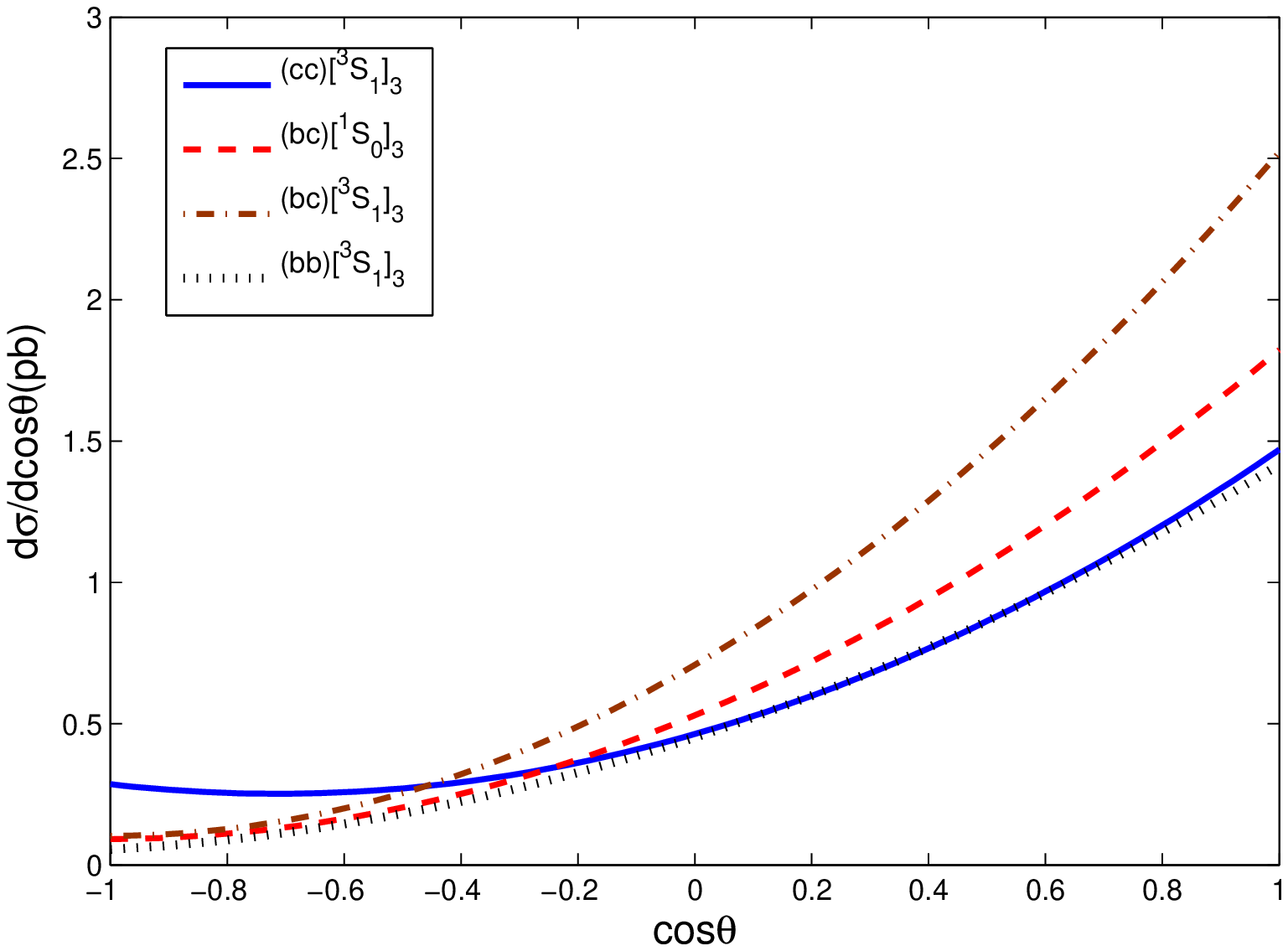}
\caption{Angle distributions of the polarized differential cross sections $d\sigma/d\cos{\theta}$ for the production of the doubly heavy baryons with $e^+_R e^-_L$  polarization states under complete QCD approach, where $\theta$ is the angle between the momenta of the final baryon and the initial electron. In order to show in one figure, for $\Xi_{bb}$, it was multiplied by a factor ten.} \label{Xicosl}
\end{figure}

\begin{figure}[!h]
\centering
\includegraphics[width=0.45\textwidth]{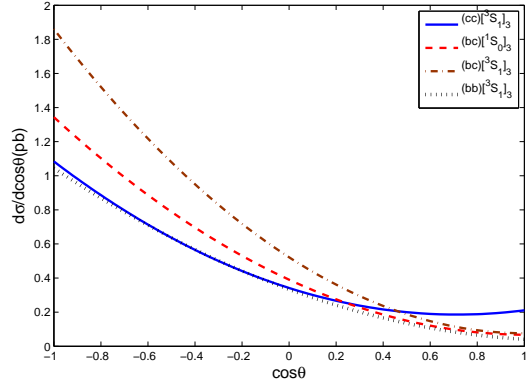}
\caption{Angle distributions of the polarized differential cross sections $d\sigma/d\cos{\theta}$ for the production of the doubly heavy baryons with $e^+_L e^-_R$  polarization states under complete QCD approach, where $\theta$ is the angle between the momenta of the final baryon and the initial electron. In order to show in one figure, for $\Xi_{bb}$, it was multiplied by a factor ten.} \label{Xicosr}
\end{figure}

As that above for $B_c$ meson production where the polarizations for incoming beams are considered, we
also compute and present the angle distributions of $\Xi_{cc}$, $\Xi_{bc}$ and $\Xi_{bb}$ production in polarizations: $e^+_R e^-_L$ and $e^+_L e^-_R$ (in the other cases the results are zero) respectively in FIGs.\ref{Xicosl},\ref{Xicosr}. From them, one can see the difference in magnitude for left-handed electron and right-handed electron easily, and it helps us understand the backward-forward asymmetry in the unpolarized differential cross sections.

\begin{figure}[!t]
\includegraphics[width=0.45\textwidth]{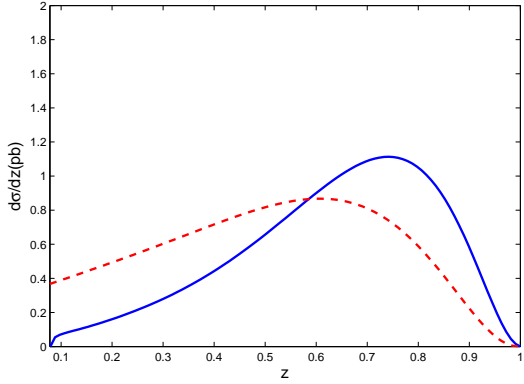}
\caption{The differential cross sections $d\sigma/dz$ in $pb$ for the production of the doubly heavy baryons $\Xi_{cc}$, where $z$ is the energy fraction of the final baryon. The solid line represents the result obtained by complete QCD approach, dashed line represents that by fragmentation approach.} \label{Xiz1}
\end{figure}

\begin{figure}[!t]
\includegraphics[width=0.45\textwidth]{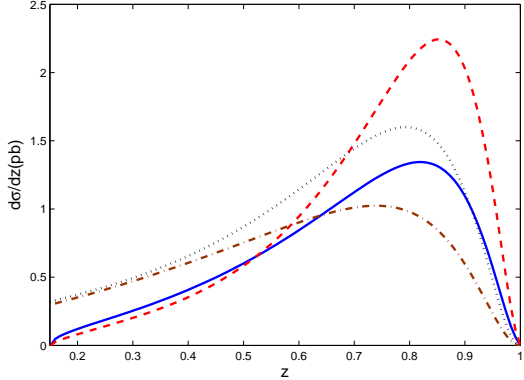}
\caption{The differential cross sections $d\sigma/dz$ in $pb$ for the production of the doubly heavy baryons $\Xi_{bc}$, where $z$ is the energy fraction of the final baryon. The solid and dashed lines represent the contributions from ${\vert\left( bc \right){\scriptsize _{\overline{\textbf{3}}}} \,, ^1S_0 \rangle}$ and ${\vert\left( bc \right){\scriptsize _{\overline{\textbf{3}}}} \,, ^3S_1 \rangle}$ obtained by complete QCD approach, dash-dot and dot lines represent the contributions from ${\vert\left( bc \right){\scriptsize _{\overline{\textbf{3}}}} \,, ^1S_0 \rangle}$ and ${\vert\left( bc \right){\scriptsize _{\overline{\textbf{3}}}} \,, ^3S_1 \rangle}$ obtained by fragmentation approach. } \label{Xiz2}
\end{figure}

\begin{figure}[!t]
\includegraphics[width=0.45\textwidth]{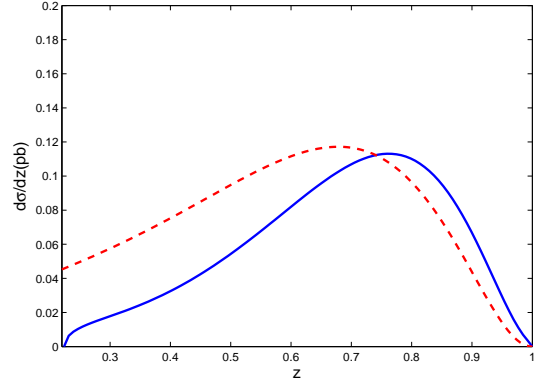}
\caption{The differential cross sections $d\sigma/dz$ in $pb$ for the production of the doubly heavy baryons $\Xi_{bb}$, where $z$ is the energy fraction of the final baryon.  The solid line represents the result obtained by complete QCD approach, dashed line represents that by fragmentation approach.} \label{Xiz3}
\end{figure}

In FIGs.\ref{Xiz1},\ref{Xiz2},\ref{Xiz3} the energy fractions of $\Xi_{cc}$, $\Xi_{bc}$ and $\Xi_{bb}$  obtained by complete QCD calculation (LO) and fragmentation approach (LL) are presented respectively. For the results by fragmentation approach, the scale of the fragmentation functions have been evolved to $m_Z$. For $\Xi_{bc}$ production, the different diquark spin-state contributions are showed respectively in the figure. According to the analysis in $B_c$ meson production, for $\Xi_{bc}$ production, the contribtuions from $c$-quark fragmentation is very small, so we only present the contribtuion from $b$-quark fragmentation for $\Xi_{bc}$ production.
From FIGs.\ref{Xiz1},\ref{Xiz2},\ref{Xiz3} one may see that the production of doubly heavy baryons at low energy fraction (small $z$) estimated by the fragmentation approach is greater than that by complete QCD approach, this is the same as we have observed in $B_c$ meson production.

\section{Disscussions and conclusion}
\label{conclusion}

We, under two approaches, estimate the production of mesons $B_c$, its excited states $B_c(B^*_c,\cdots)$ and doubly heavy baryons $\Xi_{cc}$, $\Xi_{bc}$, $\Xi_{bb}$ at a $Z$-factory with especial care on treating essential energy-scales of the production, and the results of the production
on the total cross sections are consistent with the results on branching ratios for $Z$-boson decays to $B_c$ and $B^*_c$ up-to the next leading order (NLO) calculations\cite{qiao}\footnote{We compare the results on the decay branching ratios because in Ref.\cite{qiao} only the branching ratios are available.}. Whereas since we calculate the process $e^++e^-\to B_c(B^*_c)+\bar{c}+b$ instead of the $Z$-boson decays as in Ref.\cite{qiao}, so we can obtain more properties of the production, such as the asymmetries shown in FIGs.\ref{bccos},\ref{Xicos}
and the production with polarized incoming beams shown in FIGs.\ref{bccosl},\ref{bccosr},\ref{Xicosl},\ref{Xicosr} besides the momentum fraction ($z$) distributions shown in FIGs.\ref{bcz1},\ref{bcz2},\ref{bcz3} and FIGs.\ref{Xiz1},\ref{Xiz2},\ref{Xiz3} etc. These properties
are important in studying the production of the doubly heavy hadrons. When all the experimental data are available, via comparisons of the data with the results obtained by the approaches, one may conclude which approach adopted here is better.

From the estimation here, one may expect that at a $Z$-factory with a luminosity such as $10^{34}cm^{-2}s^{-1}$ about a million of the $B_c$ mesons and doubly heavy baryons in ground-state may be produced per year, and when indirect production (to produce the excited states of them first and then
by strong or electromagnetic interaction decays to transit to the ground states accordingly) is taken into account, then a few millions of $B_c$ mesons and the doubly heavy baryons in ground-state may be produced per year. Considering the fact that the efficiency of detecting the doubly heavy hadrons is quite small\footnote{For instance, the main decays channels for experimental detecting $B_c$ meson are $B_c \to J/\Psi l^+ \nu_l$ and $B_c \to J/\Psi\pi^+$ etc, whose branch ratios for the decays are predicted to be $1\thicksim3\%$ and $0.2\thicksim0.4\%$ \cite{bcdecay}, then with cascade decay $J/\Psi\to l^+l^-$ followed, whose branching ratio is about 7\%, so the detecting efficiency would be $O(10^{-2})\times 0.07$ at most. Moreover the lifeimes of the ground doubly-heavy hadrons are $0.1\sim O(1)\; ps$, to distinguish them from background experimentally needs vertex detector for help, thus those experimental events of the doubly heavy hadrons, which have Lorentz boost not great enough, have to be cut away
due to that they cannot be distinguished from the background etc.} and the competition from LHCb  etc, one may conclude that studies of the ground states may be OK (margin), but to discovery (to study of) the excited states of the doubly heavy hadrons etc, it would be better when the $Z$-factory's luminosity is so high as $\sim 10^{35-36}cm^{-2}s^{-1}$.

Finally, in summary, several remarkable points on the production are collected as follows:

\begin{itemize}
\item To study the doubly heavy hadrons at an $e^++e^-$ collider, only when its collision energy set at $Z$-pole
    with strong resonance effect, is practicable. When the collision energy deviates from $Z$-pole, the production cross sections decrease rapidly: when the collision energy deviates from $Z$-pole by $5$ GeV, the cross sections decrease just to about 10\% of those at $Z$-pole; when
    collision energies at 180 GeV and 240 GeV, the cross sections of $B_c$ meson production become too small (in $fb$) to be observed.
\item There are obvious backward-forward asymmetry in doubly heavy hadron productions. The backward-forward asymmetry comes from the parity
    violating in these processes involving the $Z^0$ exchange. In the productions of doubly heavy hadrons at Z-factory, the main contribution comes from the $Z^0$ exchange, so the backward-forward asymmetry is very obvious. This interesting property is observable at Z-factory.
\item After summing the leading logarithm terms through the DGLAP equation, we obtain the energy distribution which is different from that obtained from the complete QCD calculation. Namely it predict more doubly heavy hadrons produced with lower energy fraction after evolving the scale to $m_Z$ in comparison with the prediction of the complete QCD prediction.
\end{itemize}

\vspace{1cm}

\vspace{4mm}

\noindent {\bf\Large Acknowledgments:} This work was supported
in part by Nature Science Foundation of China (NSFC) under Grant No. 11275243, No. 11447601.


\begin{thebibliography}{99}
\bibitem{CDF} F. Abe, et al. (CDF Collaboration), Phys. Rev. Lett.
{\bf 81}, 2432 (1998); Phys. Rev. D {\bf 58}, 112004 (1998).

\bibitem{CDF1} T. Aaltonen, et al. (CDF Collaboration). Phys. Rev. D {\bf 87}, 011101 (2013).

\bibitem{D0} V.M. Abazov, et al. (D0 Collaboration), Phys. Rev. Lett.
{\bf 101}, 012001 (2008); Phys. Rev. Lett. {\bf 102}, 092001 (2009).

\bibitem{LHCb} R. Aaij, et al . (LHCb Collaboration), Phys. Rev. Lett.
{\bf 109}, 232001 (2012); {\bf 111}, 181801 (2013); {\bf 114}, 132001 (2015); Eur. Phys. J. C. {\bf 74},
2839 (2014); J. High Energy Phys. {\bf 09} 075 (2013).

\bibitem{selex1} M. Mattson, et al. (SELEX Collaboration), Phys. Rev. Lett.
{\bf 89}, 112001 (2002).

\bibitem{selex2} M. A. Moinester, et al. (SELEX Collaboration), Czech. J.
Phys. {\bf 53}, B201 (2003).

\bibitem{selex3} A. Ocherashvili, et al. (SELEX Collaboration), Phys. Lett.
B{\bf 628}, 18 (2005).

\bibitem{eebc1} Z. Yang, X.-G. Wu, G. Chen, Q.-L. Liao and J.-W. Zhang, Phys. Rev. D{\bf 85}, 094015 (2012).

\bibitem{nrqcd} G. T. Bodwin, E. Braaten and G. P. Lepage, Phys. Rev. D {\bf 51},
1125 (1995) [Erratum-ibid. D {\bf 55}, 5853 (1997)].

\bibitem{ybook} N. Brambilla, et al. Heavy Quarkonium Physics, CERN-2005-005
20 June 2005, arXiv: hep-ph/0412158; Heavy quarkonium: progress, puzzles, and opportunities,
Eur. Phys. J. C {\bf 71}, 1534 (2011) and references therein.

\bibitem{changch} J.-P. Ma and C.-H. Chang,  Sci. China-Phys.
Mech. Astron., Preface: Z-Factory, {\bf 53}: p-1947 (2010).

\bibitem{LEP-I} R. Barate, et al. (ALAPH Collaboration) Phys. Lett. B {\bf 402} 213 (1997);
P. Abreu, et al. (DELPHI Collaboration), Phys. Lett. B {\bf 398} 207 (1997); K. Ackerstaff
{\em et al.} (OPAL Collaboration), Phys. Lett. B {\bf 420} 157 (1998).

\bibitem{pot1}  E. Eichten, K. Gottfried, T. Kinoshita, K.D. Lane and T.M. Yan, Phys.Rev. D{\bf 17}, 3090(1978); ibid. {\bf 21}, 313(E)(1980); ibid.{\bf 21}, 203(1980).

\bibitem{pot2} W. Buchm${\rm \ddot{u}}$ller and S.-H.H. Tye, Phys.Rev. D{\bf 24}, 132(1981).

\bibitem{pot3} A. Martin, Phys.Lett. B{\bf 93}, 338(1980).

\bibitem{pot4} C. Quigg and J.L. Rosner, Phys.Lett. B{\bf 71}, 153(1977).

\bibitem{pot5} Y.-Q. Chen and Y.-P. Kuang, Phys.Rev. D{\bf 46}, 1165(1992); Erratum-ibid. D{\bf 47}, 350(1993).

\bibitem{pot6} E.J. Eichten and C. Quigg, Phys.Rev. D{\bf 49}, 5845(1994).

\bibitem{zbc1} C.-H. Chang and Y.-Q. Chen, Phys. Rev. D{\bf 46}, 3845(1992).

\bibitem{fragbc1} E. Braaten, K. Cheung and T.C. Yuan, Phys. Rev. D {\bf 48}, R5049 (1993).

\bibitem{fragbc2} Y.-Q. Chen, Phys. Rev. D {\bf 48}, 5181 (1993).

\bibitem{AP} G. Altarelli and G. Parisi, Nucl.Phys. B{\bf 126}, 298(1977).

\bibitem{APsolve} R.D .Field, {\em Applications of Perturbative QCD}, Addison-Wesley(1989).

\bibitem{doublybaryon3} C.-H. Chang, C.-F.Qiao, J.-X. Wang and X.-G. Wu, Phys.Rev. D{\bf 70}, 114019(2004).

\bibitem{eeXi} J. Jiang, X.-G. Wu, Q.-L. Liao, X.-C. Zheng and Z.-Y. Fang,  Phys. Rev. D{\bf 86}, 054021 (2012).

\bibitem{PDG} K.A Olive, et al. (Particle Data Group), Chin.Phys. C{\bf 38}, 090001(2014).

\bibitem{doublybaryon4} S.P. Baranov, Phys.Rev. D{\bf 54}, 3228(1996).

\bibitem{qiao} Cong-Feng Qiao, Li-Ping Sun, Rui-Lin Zhu, JHEP {\bf 1108}, 131 (2011);
Jun Jiang, Long-Bin Chen, Cong-Feng Qiao, Phys. Rev. D{\bf 91}, 034033 (2015).

\bibitem{bcdecay} C.-H. Chang and Y.-Q. Chen, Phys. Rev. D{\bf 49}, 3399(1994).

\end{thebibliography}
\end{document}